\documentclass[a4paper,11pt]{article}
\usepackage{pos}
\usepackage{subcaption}
\usepackage[separate-uncertainty=true,multi-part-units=single]{siunitx}
\usepackage{amsmath}
\usepackage{makecell}
\usepackage{gensymb}

\title{Uncertainty study for the Galactic calibration of radio antenna arrays in astroparticle physics}
 \ShortTitle{Uncertainties in the Galactic calibration of radio arrays}

\author*[a,b]{Max Büsken}
\author[c]{Tomáš Fodran}
\author[d,e]{Tim Huege}
\affiliation[a]{Institute for Experimental Particle Physics (ETP), Karlsruhe Institute of Technology (KIT),\\ Hermann-von-Helmholtz-Platz 1, 76344 Eggenstein-Leopoldshafen, Germany}
\affiliation[b]{Instituto de Tecnologías en Detección y Astropartículas, Universidad Nacional de San Martín,\\ Av. General Paz 1555 (B1630KNA), San Martín, Buenos Aires, Argentina}
\affiliation[c]{Department of Astrophysics/IMAPP, Radboud University,\\ PO Box 9010, 6500 GL Nijmegen, The Netherlands}
\affiliation[d]{Institute for Astroparticle Physics (IAP), Karlsruhe Institute of Technology (KIT),\\ Hermann-von-Helmholtz-Platz 1, 76344 Eggenstein-Leopoldshafen, Germany}
\affiliation[e]{Astrophysical Institute, Vrije Universiteit Brussel,\\ Pleinlaan 2, 1050 Brussels, Belgium}

\emailAdd{max.buesken@kit.edu}
\emailAdd{t.fodran@science.ru.nl}
\emailAdd{tim.huege@kit.edu}

\abstract{In recent years, arrays of radio antennas operating in the MHz regime have shown great potential as detectors in astroparticle physics. In particular, they fulfill an important role in the indirect detection of ultra-high energy cosmic rays. For a proper determination of the energy scale of the primary particles, accurate absolute calibration of radio detectors is crucial. Galactic calibration – i.e., using the Galaxy-dominated radio sky as a reference source – will potentially be the standard method for this task. However, uncertainties in the strength of the Galactic radio emission lead to uncertainties in the absolute calibration of the radio detectors and, thus, in the energy scale of the cosmic-ray measurements.

To quantify these uncertainties, we present a study comparing seven sky models in the radio-frequency range of 30 to 408 MHz. By conversion to the locally visible sky, we estimate the uncertainties for the cases of the radio antenna arrays of GRAND, IceCube, LOFAR, \mbox{OVRO-LWA}, the Pierre Auger Observatory, RNO-G and SKA-low. Finally, we discuss the applicability of the Galactic calibration, for example, regarding the influence of the quiet Sun.}

\ConferenceLogo{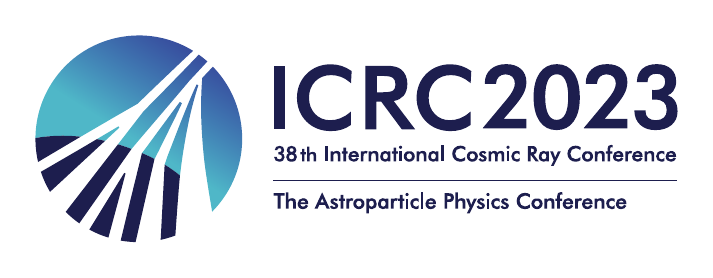}

\FullConference{%
38th International Cosmic Ray Conference (ICRC2023)\\
  26 July - 3 August, 2023\\
  Nagoya, Japan}

\begin{document}
\maketitle

\section{Introduction}

Cosmic particles, like ultra-high energy cosmic rays and neutrinos, are fascinating messengers from outside our Galaxy. Detecting those particles at the highest energies is only possible with indirect methods, for example by observing cascades of secondary particles induced by the cosmic primaries when interacting with a medium, e.g.\ the atmosphere or ice. A particular method that gained a lot of maturity in the past years is the detection of radio signals emitted by those particle cascades with antenna arrays \cite{Huege_2016, Schroeder_2017}. As for any other detection technique, radio detectors need to be accurately calibrated, both in their directional response and on an absolute scale. One typical calibration method for this task is a dedicated campaign with an artificial reference emitter mounted on a drone that is flown around an antenna station. This method is suitable for determining the antenna's directional response but exhibits drawbacks about its practicability (especially for long-term calibration) and can suffer from considerable uncertainties on the absolute signal strength of the reference emitter \cite{Mulrey_2019}.

A promising method, which facilitates long-term calibration and does not require dedicated campaigns, is the Galactic calibration. The Galactic radio emission is the dominant background signal at low frequencies in locations where there is not much anthropogenic noise \cite{ITU_Noise}. This fact can be exploited to calibrate radio detectors by comparing measured background signal to a predicted signal which can be estimated with models of the Galactic radio emission and an accurate detector description. Galactic calibration studies were already conducted at LOFAR \cite{Mulrey_2019} and the Pierre Auger Observatory \cite{Fodran_2021, deAlmeida_2023_2}.

In this work, building up on Ref.\ \cite{Buesken_2022}, uncertainties are estimated on how accurately the Galactic signal can be predicted. That way, the capabilities of using the Galactic calibration for setting an absolute energy scale for radio detectors are probed. We perform this uncertainty estimation by conducting a comparison of radio sky models in the frequency range from 30 to \SI{408}{MHz} and by discussing influences from other natural sources of radio emission.

\section{Predicting the Galactic radio emission}
Multiple radio sky models were developed in the last years that can generate full-sky maps of the radio emission. They consist of physical descriptions or mathematical models of the radio sky and are either tuned to reference surveys or directly built from these. While the reference surveys are set at specific frequencies, the output maps generated from the models can be at any frequency within the range of the respective model.

\begin{figure*}
  \begin{subfigure}[c]{0.48\hsize}
    \center
    \includegraphics[width=1.\hsize]{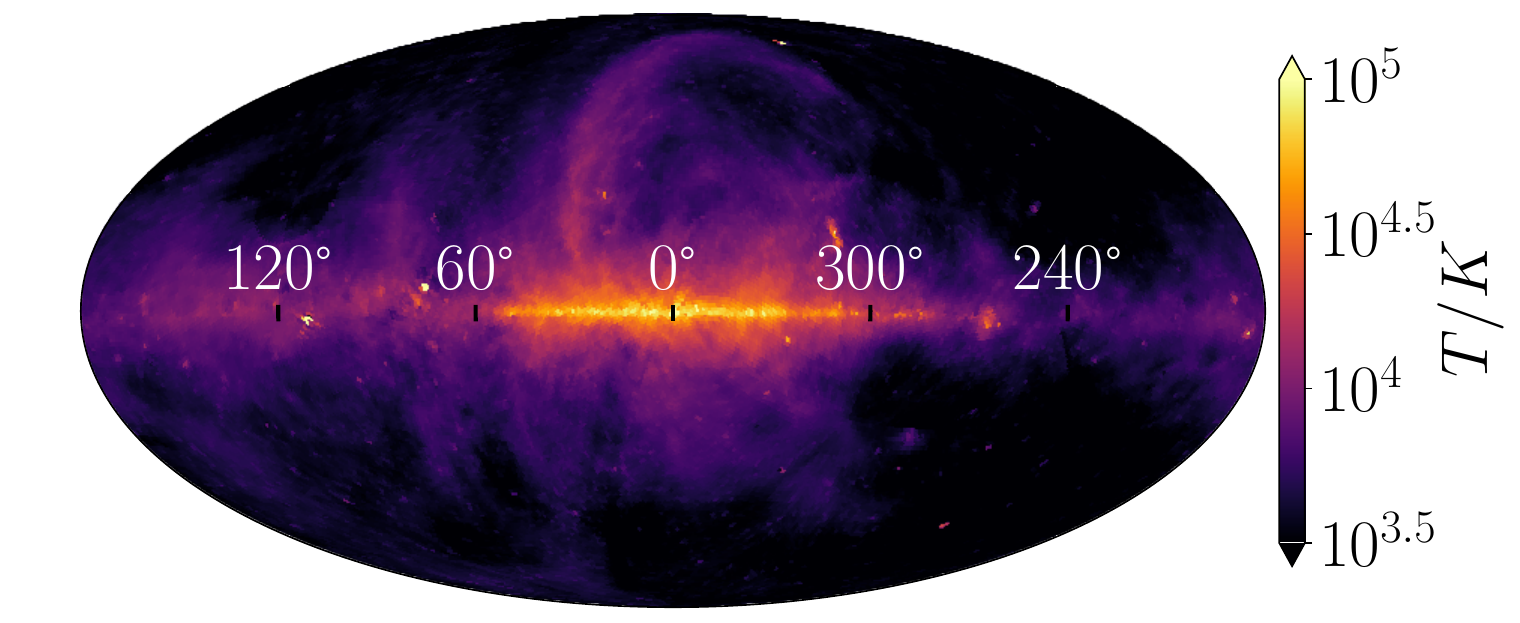}
    \subcaption{\;\;\;\;\;\;}
  \end{subfigure}
  \begin{subfigure}[c]{0.48\hsize}
    \center
    \includegraphics[width=1.\hsize]{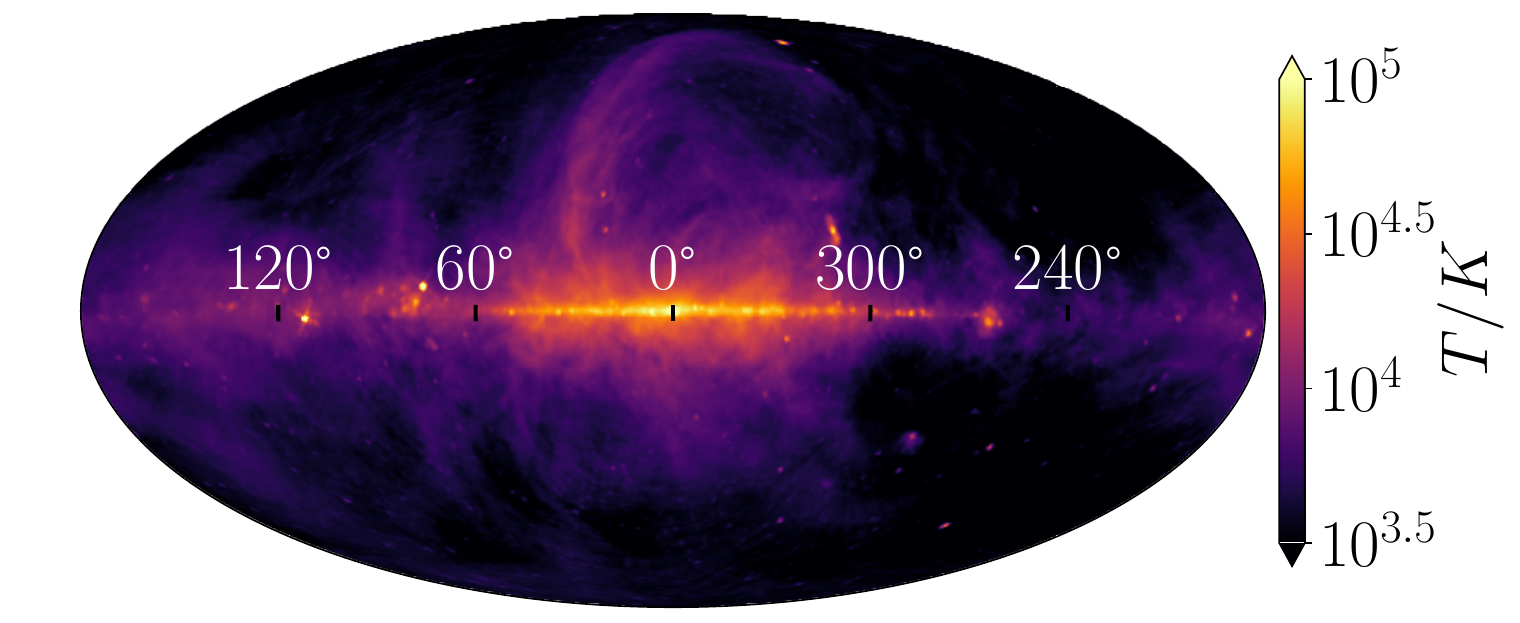}
    \subcaption{\;\;\;\;\;\;}
  \end{subfigure}
  \\
  \begin{subfigure}[c]{0.48\hsize}
    \center
    \includegraphics[width=1.\hsize]{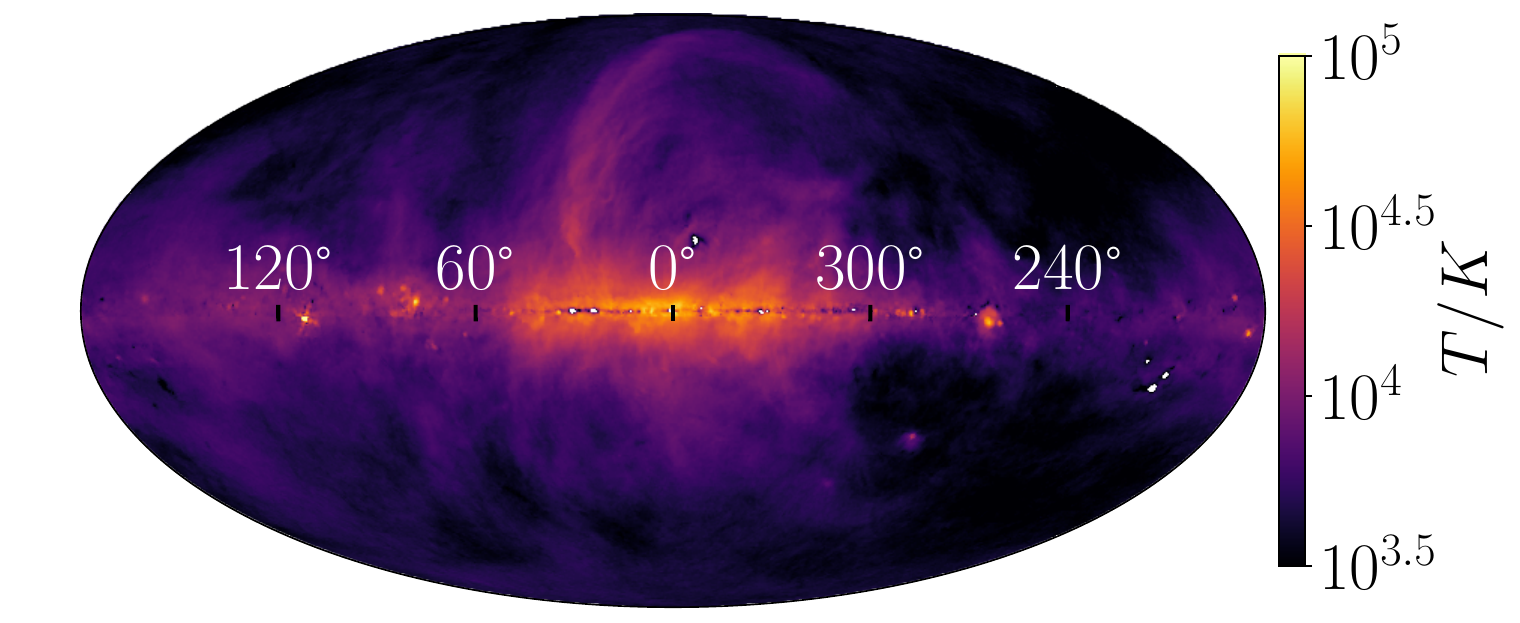}
    \subcaption{\;\;\;\;\;\;}
  \end{subfigure}
  \begin{subfigure}[c]{0.48\hsize}
    \center
    \includegraphics[width=1.\hsize]{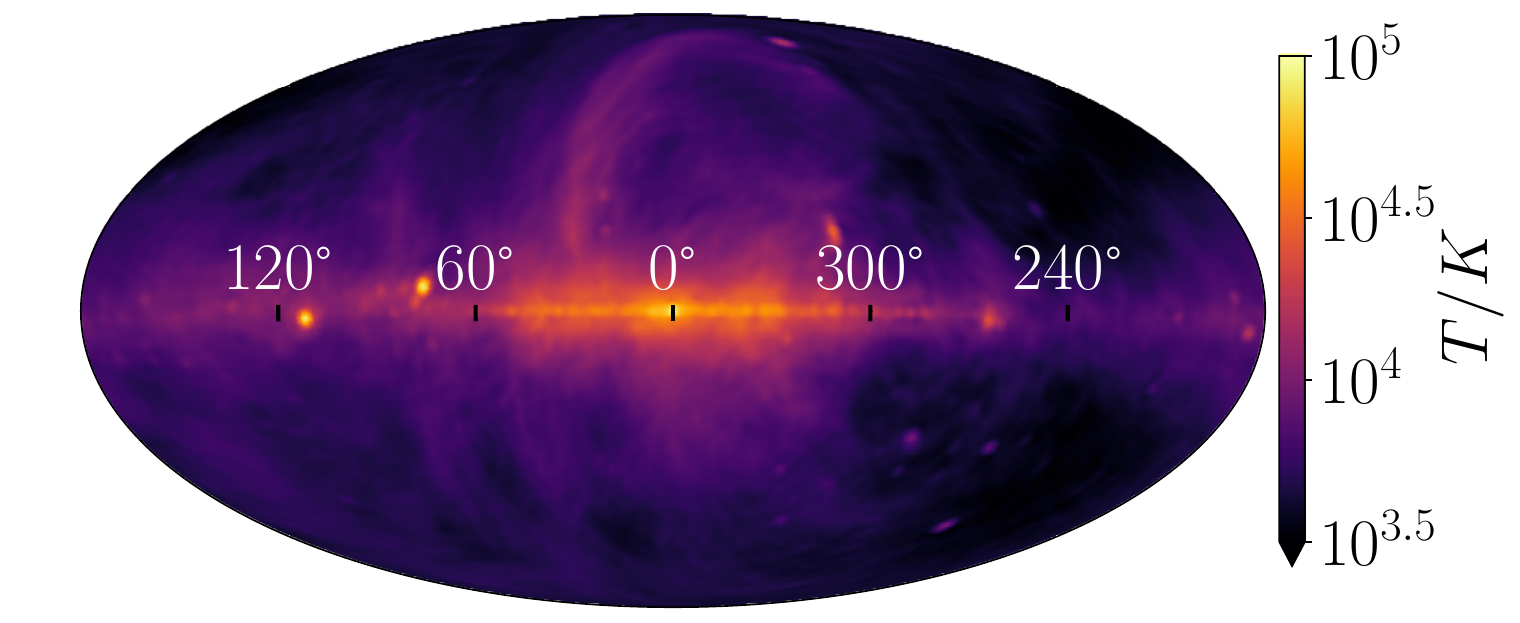}
    \subcaption{\;\;\;\;\;\;}
  \end{subfigure}
  \\
  \begin{subfigure}[c]{0.48\hsize}
    \center
    \includegraphics[width=1.\hsize]{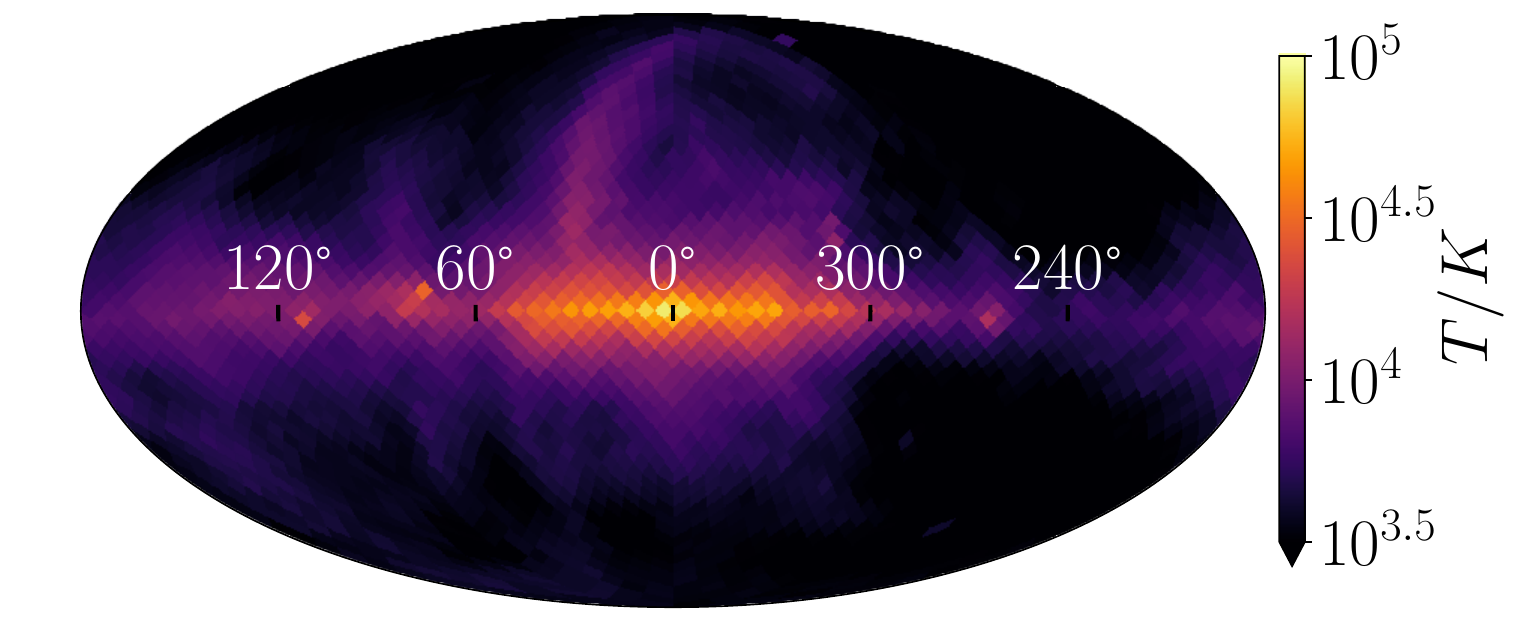}
    \subcaption{\;\;\;\;\;\;}
  \end{subfigure}
  \begin{subfigure}[c]{0.48\hsize}
    \center
    \includegraphics[width=1.\hsize]{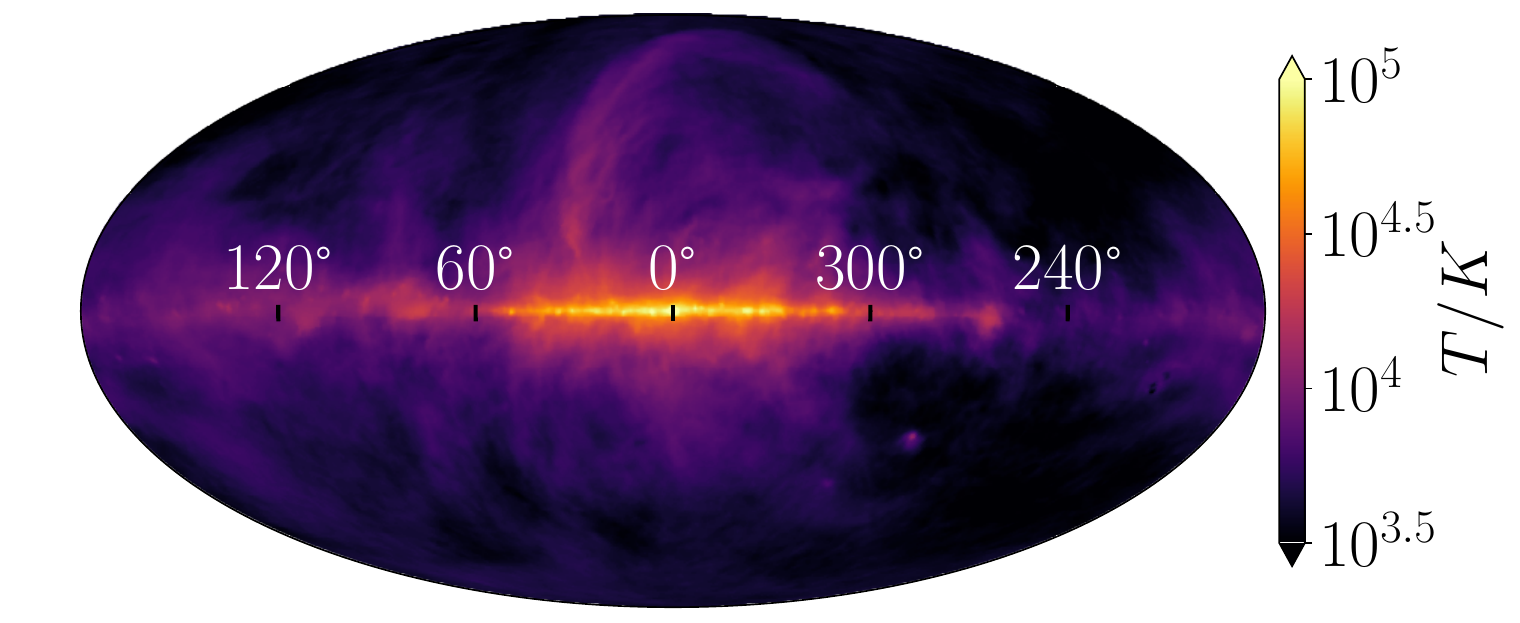}
    \subcaption{\;\;\;\;\;\;}
  \end{subfigure}
  \\
  \hspace*{\fill}
  \begin{subfigure}[c]{0.48\hsize}
    \center
    \includegraphics[width=1.\hsize]{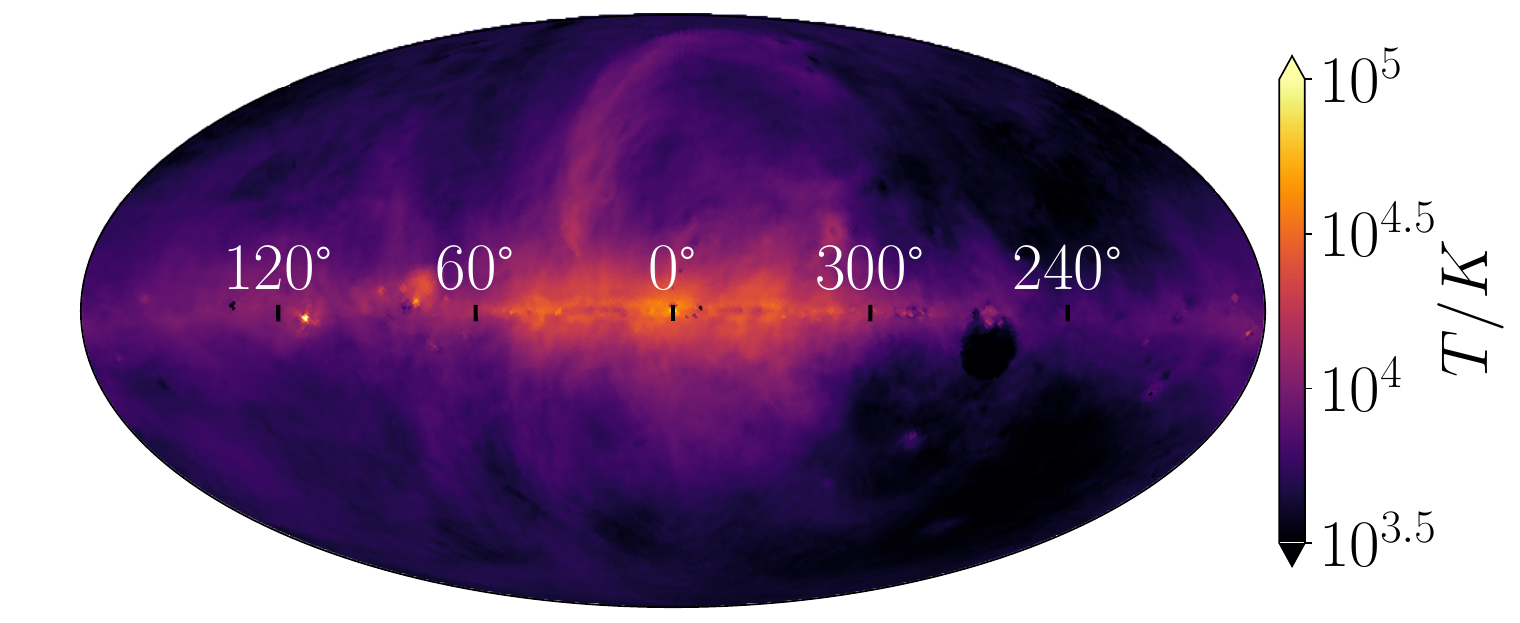}
    \subcaption{\;\;\;\;\;\;}
  \end{subfigure}
  \hspace*{\fill}
  \caption{Maps of the radio sky at \SI{50}{MHz} generated by each of the models in Galactic coordinates: (a) LFmap, (b) GSM, (c) GSM16, (d) LFSM, (e) GMOSS, (f) SSM, (g) ULSA.}
  \label{fig:example_maps}
\end{figure*}

In this study, we use seven publicly available radio sky models, LFmap, GSM (2008), GSM (2016), LFSM, GMOSS, SSM and ULSA. A summary of the core modeling approaches and the reference maps used in each of the models is given in Tab.\ \ref{tab:sky_models}. Maps of the radio sky are typically presented in the brightness temperature $T$, which in good approximation scales linearly with the brightness $I_\nu$ at low frequencies $\nu$ as $T = \frac{c^2}{2k_{\text{B}} \nu^2}I_{\nu}$. Exemplarily, maps generated with each of the studied sky models at \SI{50}{MHz} are shown in Fig.\ \ref{fig:example_maps}. The Galaxy can be seen to dominate the radio emission, while the brightness drops off toward the Galactic poles. There are small differences visible between the models. Two reasons for these differences are the disparate modeling approaches and the different sets of reference maps.

\begin{table}
	\centering
	\begin{tabular}{c | c | c | c}
	Sky model & Modeling approach & \makecell{Reference maps at\\ low frequencies / MHz} & Refs. \\ \hline
      LFmap  & Spectral scaling                             & 22, 408 & \cite{LFmap} \\
      GSM (2008)  & Principal component analysis            & 10, 22, 45, 408 & \cite{de_Oliveira_Costa_2008} \\
      GSM (2016)  & Principal component analysis            & 10, 22, 45, 85, 150, 408 & \cite{Zheng_2016} \\
      LFSM  & Principal component analysis                  & \makecell{10, 22, 40, 45, 50, 60,\\ 70, 80, 408} & \cite{Dowell_2017} \\
      GMOSS  & Physical emission model                      & 22*, 45*, 150, 408 & \cite{GMOSS} \\
      SSM  & Spectral scaling + point sources               & 10, 22, 45, 408 & \cite{SSM} \\
      ULSA  & Physical emission model + spectral scaling    & \makecell{35, 38, 40, 45, 50, 60,\\ 70, 74, 80, 408} & \cite{ULSA} \\
	\end{tabular}
	\caption{\label{tab:sky_models}Summary of the radio sky models used in this study. The asterisk denotes maps that were generated with GSM (2008) and are used in the respective model as reference maps.}
\end{table}

\section{Comparison of model predictions}

In order to estimate the accuracy to which the absolute strength of the Galactic radio emission can be predicted, we determine the level of agreement between the output maps generated by the sky models. The level of agreement is our estimator for the uncertainty of the absolute temperature scale, which could be applied in Galactic calibration studies of radio arrays. We believe that the considered set of models is exhaustive in the sense that it adequately reflects the current knowledge of the the Galactic radio emission.

We conduct the model comparison by calculating the average sky temperature from a full-sky map in Galactic coordinates (longitude $\ell$, latitude $b$) as 

\begin{equation}
\label{eq:Tsky_average}
\bar{T}(\nu) = \frac{1}{4\pi} \int_{-\pi}^{\pi} \mathrm{d} \ell \int_{\frac{-\pi}{2}}^{\frac{\pi}{2}} \mathrm{d} b \cos{(b)}\; T(\nu; \ell, b).
\end{equation}

\begin{figure}
    \centering
        \includegraphics[width=0.7\textwidth]{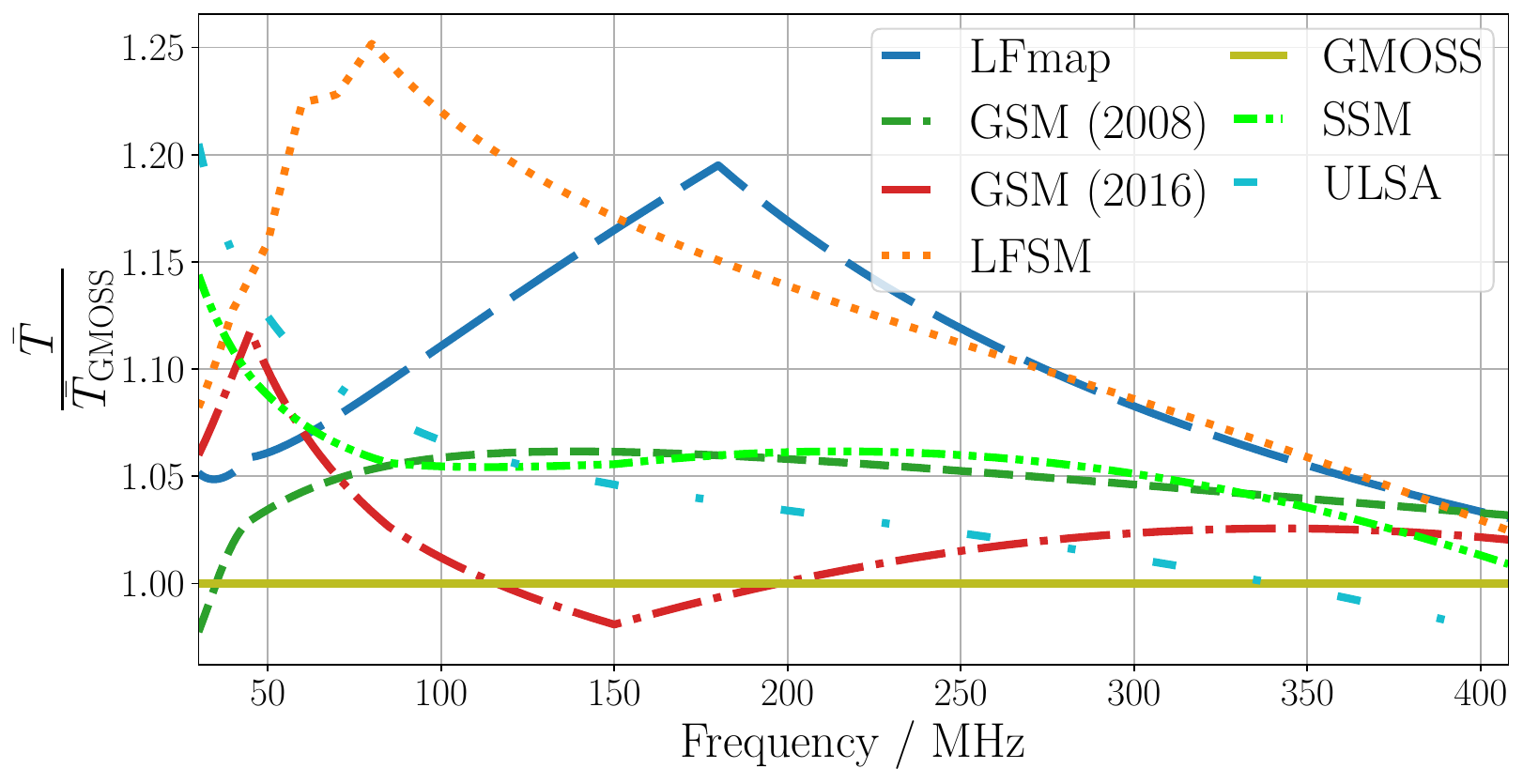}
	\caption{Ratio of the average sky temperature of the radio sky between each of the sky models and GMOSS, plotted as a function of the frequency. }
	\label{fig:All_models_normed_log}
\end{figure}

The average sky temperature as a function of the frequency from \SIrange{30}{408}{MHz} is plotted in Fig.\ \ref{fig:All_models_normed_log} for each of the sky models, normalized to the GMOSS result. GMOSS predicts the faintest radio skies for large parts of the considered frequency range. Deviations between the models are largest at the lower frequencies with up to \SI{25}{\%} between GMOSS and LFSM around \SI{80}{MHz} but decrease towards \SI{408}{MHz}.

For determining a global estimate of the absolute scale uncertainty of the Galactic radio emission in the studied frequency range, we define the following ratio:

\begin{equation}
\label{eq:ratio}
r_{\text{m}_1,\text{m}_2} = 2 \frac{\int_{30\text{MHz}}^{408\text{MHz}} \bar{T}_{\text{m}_1}(\nu) - \bar{T}_{\text{m}_2}(\nu) \; \text{d} \nu }{\int_{30\text{MHz}}^{408\text{MHz}} \bar{T}_{\text{m}_1}(\nu) + \bar{T}_{\text{m}_2}(\nu) \; \text{d} \nu } \text{.}
\end{equation}

Here, $\text{m}_1$ and $\text{m}_2$ represent any combination of sky models. We suggest the maximum value of this ratio as an uncertainty estimate. This maximum is \SI{14.3}{\%} for the ratio between GMOSS and LFSM.

Next, we investigate differences between the models as a function of the geographical latitude of an observer on earth who would not see the full sky but only part of it. We convert the generated full-sky maps to the horizontal coordinate system with altitude a and azimuth $\alpha$ at a geographic latitude $\ell$ and local sidereal time (LST) $t_\text{LST}$. We average the observed sky temperature $\bar{T}_\text{local}$ over the local sky and LST. Then, we integrate over the frequency range $\nu_\text{low}$, $\nu_\text{high}$ and obtain

\begin{equation}
\label{eq:local_sky_t}
\mathcal{T} (\ell) = \frac{1}{2\pi} \frac{1}{\SI{24}{h}} \int_{\nu_\text{low}}^{\nu_\text{high}} \text{d} \nu  \int_{\SI{0}{h}}^{\SI{24}{h}} \text{d} t_{\text{LST}} \int_{0}^{\pi} \text{d} a \int_{\frac{-\pi}{2}}^{\frac{\pi}{2}} \text{d} \alpha \cos{(a)}\; T(\nu, \ell, t_{\text{LST}}; a, \alpha).
\end{equation}

\begin{figure}
    \centering
        \includegraphics[width=1.0\textwidth]{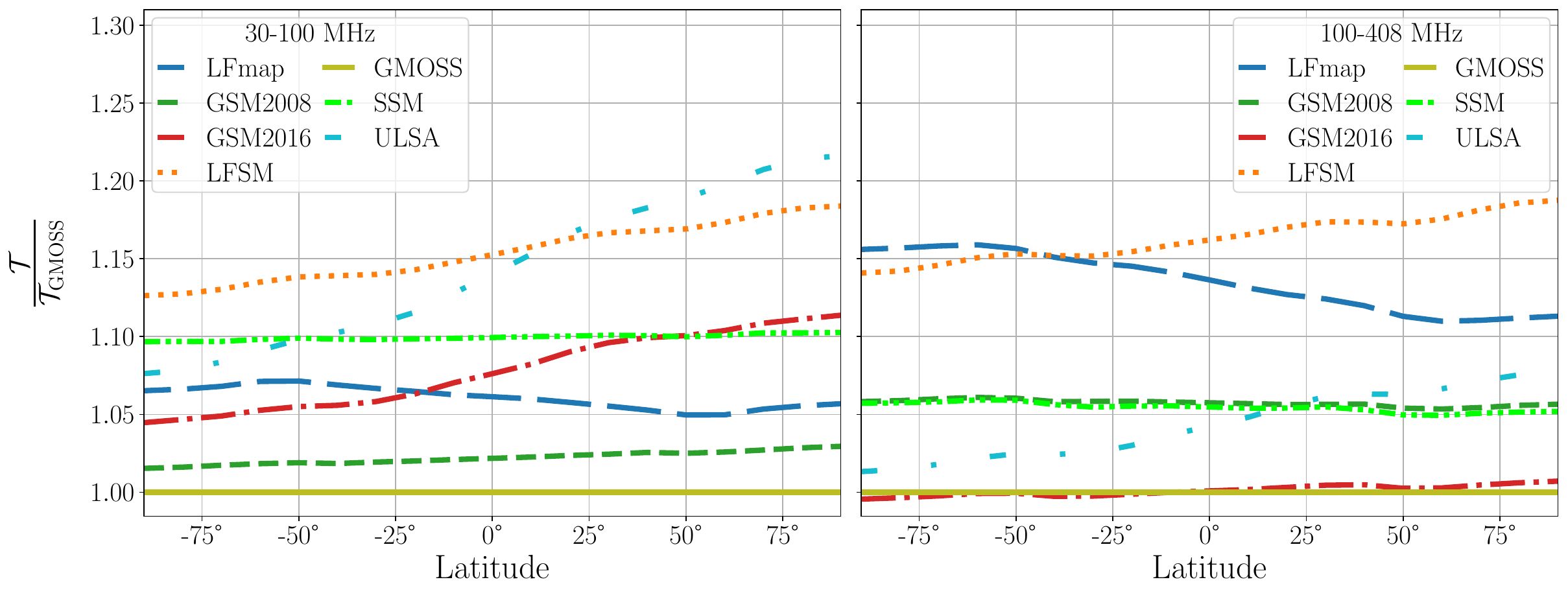}
	\caption{}
	\label{fig:All_models_local_lat_combined_normalized}
\end{figure}

We plot $\mathcal{T}$ for each model, normalized to GMOSS, as a function of the latitude in Fig.\ \ref{fig:All_models_local_lat_combined_normalized} for the two frequency ranges from 30 to \SI{100}{MHz} and from 100 to \SI{408}{MHz} on the left and right, respectively. The ordering of the models from brightest to faintest predictions is clearly different with respect to two considered frequency regimes, although the size of the general spread between the set of models does not change. The latter, however, is dependent on the geographic latitude. The models tend to agree better at low latitudes where the Galactic center rises high in the sky, compared to high latitudes where the Galactic center stays lower in the sky.

If the Galactic calibration is to be applied to a particular radio antenna array, a dedicated analysis is necessary that takes into account the geographical position of the array, the gain pattern of the used antenna and the response of the electronics chain of the detector. We cannot provide this level of detail in this study but perform comparisons with simplifying assumptions for the radio arrays GRAND \cite{GRAND_Design}, IceCube \cite{IceCube_SurfaceArray_Development}, the LOFAR low- and high-band antennas \cite{Schellart_2013, Nelles_2015}, \mbox{OVRO-LWA} \cite{Monroe_2020}, the Pierre Auger Observatory \cite{AugerPrime_Radio}, \mbox{RNO-G} \cite{RNO-G_Design} and \mbox{SKA-low} \cite{Buitink_2021}. We calculate ratios for each array in the same way as in Eq.\ \ref{eq:ratio} but use an average sky temperature $\bar{T}_{\text{exp}}$ that is calculated from the local sky seen by the array at geographical latitude $\ell_\text{exp}$ and limit the angular integration to elevations larger than \SI{15}{\degree} above the horizon to mimic typical antenna gain patterns. Analogue to the previous paragraph, we average over \SI{24}{h} of LST. Also, we adapt the frequency band in the spectral integral to the operational frequency band of the array ($\nu_\text{exp, lower}$, $\nu_\text{exp, upper}$) to obtain 

\begin{equation}
\mathcal{T_\text{exp}} (\ell_\text{exp}) = \int_{\nu_\text{exp, lower}}^{\nu_\text{exp, upper}} \bar{T}_{\text{exp}}(\nu, \ell_\text{exp}) \; \text{d} \nu ,
\end{equation}

which we use to calculate the ratio

\begin{equation}
\label{eq:ratio_exp}
r_{\text{exp; m}_1\text{, m}_2} = 2 \frac{ \mathcal{T}_{\text{exp, m}_1} (\ell_\text{exp}) - \mathcal{T}_{\text{exp, m}_2} (\ell_\text{exp}) }{ \mathcal{T}_{\text{exp, m}_1} (\ell_\text{exp}) + \mathcal{T}_{\text{exp, m}_2} (\ell_\text{exp}) }.
\end{equation}

With that, there is a combination of models $\text{m}_1$ and $\text{m}_2$ that results in the maximum ratio for each experiment. These maxima are listed in Tab.\ \ref{tab:experiments_comparison}. They exhibit a clear dependence of the accuracy of predicting the Galactic radio emission on the parameters of the array and range between \SI{11.7}{\%} and \SI{21.5}{\%} for Auger and GRAND, respectively. These determined uncertainty estimates refer to temperature. When propagating them onto the energy scale of a radio detector the uncertainties decrease to about half of those values.

\begin{table}
    \centering
    \begin{tabular}{r | r | r | c | c}
    Observatory & $l_\text{exp}$ & Frequency band & $\max(r_{\text{exp; m}_1\text{, m}_2})$ & Corresponding sky models \\ \hline
    RNO-G  & \SI{72.58}{\degree}  & 100 to 408 MHz & \SI{17.1}{\%}   & LFSM / GSM16   \\ 
    LOFAR low    & \SI{52.91}{\degree}  & 30 to 80 MHz  & \SI{19.8}{\%}   & ULSA / GMOSS  \\
    LOFAR high    & \SI{52.91}{\degree}  & 110 to 190 MHz  & \SI{18.4}{\%}   & LFSM / GSM16  \\
    GRAND  & \SI{42.93}{\degree}  & 50 to 200 MHz  & \SI{21.5}{\%}   & LFSM / GMOSS  \\
    OVRO-LWA    & \SI{37.23}{\degree} & 30 to 80 MHz   & \SI{19.3}{\%}   & ULSA / GMOSS  \\
    SKA-low  & \SI{-26.70}{\degree} & 50 to 350 MHz  & \SI{15.1}{\%}   & LFSM / GMOSS  \\
    Auger    & \SI{-35.21}{\degree} & 30 to 80 MHz   & \SI{11.7}{\%}   & LFSM / GMOSS  \\
    IceCube  & \SI{-90.0}{\degree}  & 70 to 350 MHz & \SI{20.3}{\%}   & LFSM / GMOSS  \\
    \end{tabular}
    \caption{\label{tab:experiments_comparison} Overview of selected radio arrays in astroparticle physics and their parameters. As estimators for the uncertainties on the accuracy of predicting the Galactic radio emission on an absolute scale the maximum ratio from Eq.\ \ref{eq:ratio_exp} and the combination of the brightest and faintest model in that case are listed.}
\end{table}

\section{Other confounding factors in the Galactic calibration}

\begin{figure}
    \centering
	\includegraphics[width=0.71\textwidth]{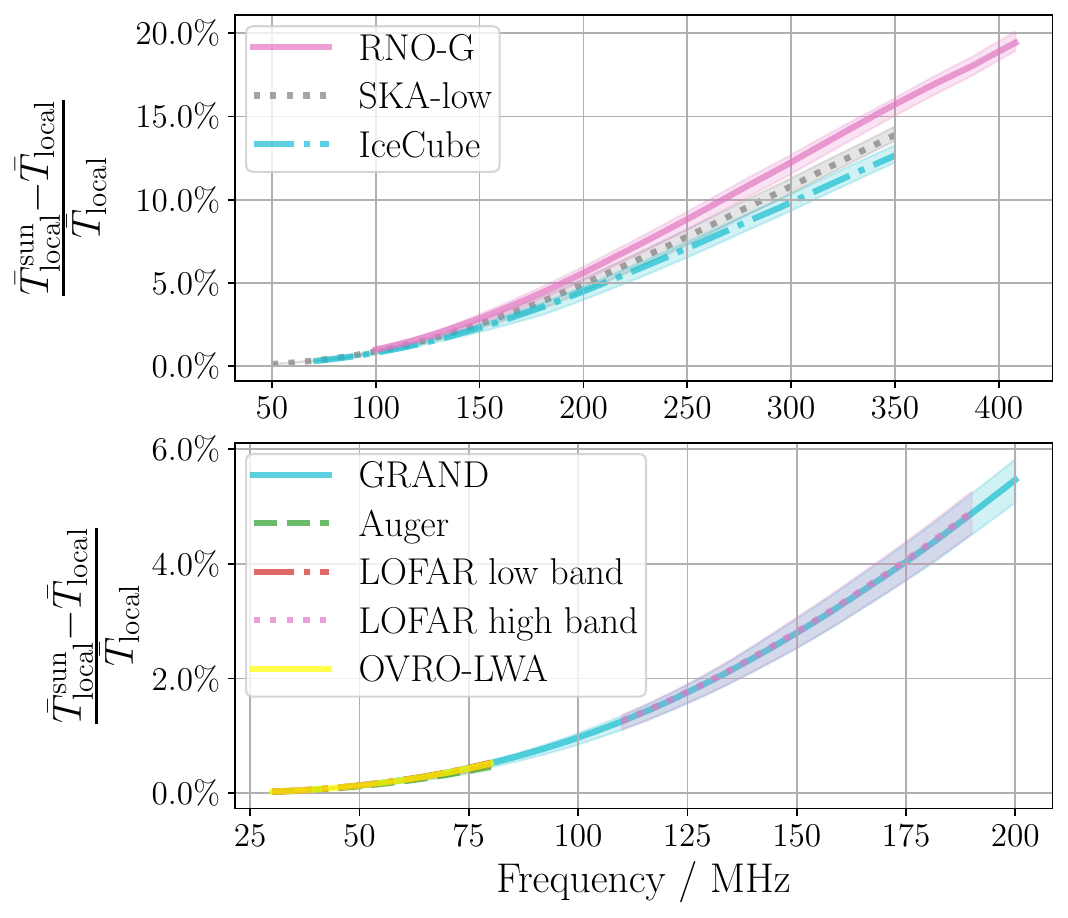}
	\caption{Influence of the quiet sun on the average temperature of the radio sky as a function of the frequency for each of the radio arrays. In the top plot, arrays that measure at higher frequencies are included, while the bottom plot contains arrays with lower observation frequencies. The influence is given as the relative difference in the average sky temperature when superimposing the radio emission of the sun onto the sky maps. Colored bands reflect the scatter resulting from the choice of sky model for generating the sky maps, while the lines represent the average from all models.}
	\label{fig:All_models_local_sun}
\end{figure}

Applying the Galactic calibration to the radio detectors of an antenna array does not only require an estimation of the absolute accuracy of the Galactic emission prediction. There are other confounding factors that need to be considered, for example the sun. In its quiet state, the sun constantly emits radio waves in a well measured brightness spectrum \cite{Kraus_RadioAstronomy,Zhang_2022}. This emission will be picked up by radio antennas at daytime and make an additional contribution to the detector background. In order to estimate the relative emission strength of the quiet sun compared to the Galactic signal, we lay a circle with the size of the sun (\SI{0.5}{\degree} angular diameter) and brightness temperature taken from Ref.\ \cite{Zhang_2022} onto a sky map as seen by a radio array, generated by a model for a given frequency. We calculate the relative difference of the average sky temperature for the addition of this model of the quiet sun and plot it in Fig.\ \ref{fig:All_models_local_sun} as a function of the frequency for the same arrays as in Tab.\ \ref{tab:experiments_comparison}. The relative difference increases with frequency. While it is at the sub-percent level for Auger and the LOFAR low-band configuration, it becomes very relevant for arrays like \mbox{RNO-G}, \mbox{SKA-low} and IceCube, reaching \SI{20}{\%} around \SI{400}{MHz}.

Other noteworthy influences may arise from the active state of the sun, Jupiter and the ionosphere. The first two have discontinuous radio emission, also within the frequency range studied in this work \cite{Kraus_RadioAstronomy, Zarka_2005, Cecconi_2012}. However, the discontinuous character makes a qualitative analysis of the influence on the Galactic calibration difficult and requires dedicated investigations. The ionosphere becomes increasingly reflective to radio waves below \SI{20}{MHz} causing emission from distant sources (e.g.\ thunderstorms) to enter a radio detector \cite{Allan_1971}. Radio arrays in astroparticle physics operate above \SI{30}{MHz} to avoid those strong ionospheric influences.

\section{Conclusions}

We present a study of uncertainties in the Galactic calibration of astroparticle radio arrays based on a comparison of radio sky models. In this comparison, we determine the level of agreement between the models, which we suggest to use as an estimator for the uncertainty of how accurately the Galactic radio emission can be predicted on an absolute scale. We find a global estimator of \SI{14.3}{\%} uncertainty on the sky brightness temperature. The energy scale of particles probed with astroparticle radio arrays scales approximately with the square root of the signal power. Since the received power is the quantity calibrated in the Galactic calibration, the found uncertainty on the energy scale is only half of the value quoted for sky temperature, i.e.\ around \SI{7}{\%}. These uncertainties show that the Galactic calibration is very competitive in accuracy with other existing calibration methods.

We expand the comparison with toy models of GRAND, IceCube, LOFAR, \mbox{OVRO-LWA}, the Pierre Auger Observatory, \mbox{RNO-G} and \mbox{SKA-low}, resulting in array-specific uncertainty estimates between \SI{11.7}{\%} and \SI{21.5}{\%}.

Finally we discuss other sources of radio emission that may influence the Galactic calibration, namely the Sun, Jupiter and the ionosphere. In a quantitative estimation, we find that the quiet sun only has a small effect at the lowest frequencies, while for arrays operating at higher frequencies, its contribution is significant. Therefore, the solar contribution may need to be removed from the background signal before calibrating those arrays, if not restricting the Galactic calibration to nighttime only.

\setlength{\bibsep}{0pt plus 0.3ex}
\bibliographystyle{MyJHEP}
\bibliography{References}

\end{document}